\documentclass[12pt,preprint]{aastex}
\usepackage{mathrsfs}

\usepackage{amsmath}

\usepackage{graphics}

\newcommand{\bd}{\begin{displaymath}}
\newcommand{\ed}{\end{displaymath}}

\slugcomment{Received 2015 November 22; accepted 2016 June 2}
\shorttitle{On the thermal line emission from the outflows in ULXs}

\begin{document}

\title{On the thermal line emission from the outflows in
ultraluminous X-ray sources}

\author{Ya-Di Xu\altaffilmark{1}, and Xinwu Cao \altaffilmark{2,3}}
\altaffiltext{1}{Department of Physics and Astronomy, Shanghai Jiao
Tong University, 800 Dongchuan Road, Shanghai 200240, China
ydxu@sjtu.edu.cn} \altaffiltext{2}{SHAO-XMU Joint Center for
Astrophysics, Shanghai Astronomical Observatory, Chinese Academy of
Sciences, 80 Nandan Road, Shanghai, 200030, China; cxw@shao.ac.cn}
\altaffiltext{3}{Key Laboratory of Radio Astronomy, Chinese Academy
of Sciences, 210008 Nanjing, China}

\begin{abstract}
The atomic features in the X-ray spectra of the ultraluminous X-ray
source (ULX) may be associated with the outflow
\citep*[][]{2015MNRAS.454.3134M}, which may provide a way to explore
the physics of the ULXs. We construct a conical outflow model, and
calculate the thermal X-ray Fe emission lines from the outflows. Our
results show that thermal line luminosity decreases with increasing
outflow velocity or/and opening angle of the outflow for a fixed
kinetic power of the outflows. Assuming the kinetic power of the
outflows to be comparable with the accretion power in the ULXs, we
find that the equivalent width can be several eV for the thermal
X-ray Fe emission line from the outflows in the ULXs with stellar
mass black holes. The thermal line luminosity is proportional to
$1/M_{\rm bh}$ ($M_{\rm bh}$ is the black hole mass of the ULX). The
equivalent width decreases with the black hole mass, which implies
that the Fe line emission from the outflows can hardly be detected
if the ULXs contain intermediate mass black holes. Our results
suggest that the thermal X-ray Fe line emission should be
preferentially be detected in the ULXs with high kinetic power
slowly moving outflows from the accretion disks surrounding stellar
mass black holes/neutron stars. The recent observed X-ray atomic
features of the outflows in the ULX may imply that it contains a
stellar mass black hole/neutron star \citep*[][]{2016.1604.08593}.
\end{abstract}


\section{Introduction}

The nature of ultraluminous X-ray sources (ULXs) is still unclear
mainly due to the lack of dynamical mass determination of their
central sources, though progress has been achieved in recent years
\citep*[][]{2013Natur.503..500L,2014Natur.514..198M}. This makes it
difficult to resolve the accretion mode in ULXs \citep*[see][for a
review, and the references theirin]{2016arXiv160105971F}. The ULXs
may either be powered by accretion on to stellar mass black holes at
high rates, or associated with accretion at $\sim$Eddington rates on
to intermediate mass black holes with $M_{\rm bh}\ga 100 {\rm
M}_\odot$. There is evidence of accretion on to stellar mass black
holes in some individual ULXs
\citep*[][]{2013Natur.503..500L,2014Natur.514..198M}, though the
nature of the compact objects in most of the ULXs is somewhat
ambiguous.

The super-Eddington accretion on to a stellar mass black hole in the
ULX can be an analogy with SS433, a well known super critical
accreting source in our galaxy
\citep*[e.g.,][]{2001IAUS..205..268F,2004ASPRv..12....1F}. The
Doppler shifted X-ray emission lines have been observed in the X-ray
spectra of SS433, which are supposed to be emitted from the jets
\citep*[][]{1986MNRAS.222..261W,1994PASJ...46L.147K,1996PASJ...48..619K,2013ApJ...775...75M}.
The X-ray line emission from the jets implies a baryonic component
in the jets \citep*[][]{2013Natur.504..260D}, which strongly
supports the Blandford-Payne jet formation mechanism at least for
this source \citep*[][]{1982MNRAS.199..883B}. A radio-X-ray
correlation in Cygnus X-1 is found to extend to the high/soft state
only if its hard X-ray emission is considered (after subtracting the
blackbody component emitted from the thin disk)
\citep*[][]{2011MNRAS.416.1324Z}. A similar feature has been found
in active galactic nuclei \citep*[][]{2013ApJ...770...31W}. These
imply that the jet formation may probably be related to the hot
coronae above the thin disks
\citep*[][]{2002MNRAS.332..165M,2004ApJ...613..716C,2011MNRAS.416.1324Z,2013ApJ...770...31W}.
The calculations of magnetically accelerated outflows show that hot
gas (probably in the corona) is necessary for launching an outflow
from the radiation-pressure-dominated accretion disk
\citep*[][]{2014ApJ...783...51C}. If this is the case, a fraction of
the hot ions and electrons in the corona may be driven into the
jets/outflows by the magnetic field of the disk. The gas in the
jets/outflows with suitable temperature may emit X-ray lines mainly
due to the re-combination of the ions and elections in the jets.

The observed Doppler-shifted X-ray emission lines provide useful
constraints on the properties of the jets in SS433
\citep*[][]{1996PASJ...48..619K,2013ApJ...775...75M}. It was claimed
that the Doppler shifted X-ray lines had been detected in the X-ray
binary 4U~1630$-$47
\citep*[][]{2000ApJ...529..952C,2013Natur.504..260D}, though it was
not confirmed \citep*[][]{2014ApJ...784L...5N}. The X-ray
observation of the ULX NGC 1313 X-1 provides evidence of soft X-ray
atomic features associated with the winds or outflows
\citep*[][]{2015MNRAS.454.3134M,2016.1604.08593}. The thermal X-ray
line emission from the outflows may provide useful clues on the
nature of ULXs.

In this work, we calculate the thermal X-ray line emission from the
outflows in ULXs, and show how the physical properties of the
outflows are related to the X-ray line emission. We describe the
outflow model in Section \ref{model}, and the calculations of the
thermal X-ray line emission in Section \ref{therm_line}. Sections
\ref{results} and \ref{discussion} contain the results and
discussion.

\section{Outflow model}\label{model}

The gas in the corona above a thin disk is very hot. The temperature
of the ions in the corona is nearly virialized ($\sim 10^{11-12}$K
in the inner region of the disk), which is much higher than the
electron temperature ($\sim 10^9$K)
\citep*[e.g.,][]{2002MNRAS.332..165M,2003ApJ...587..571L,2009MNRAS.394..207C}.
A small fraction of the hot gas in the corona may probably
accelerated into the outflows by the magnetic field co-rotating with
the gas in the disk or/and the radiation force of the disk
\citep*[][]{2014ApJ...783...51C}. The calculations of the outflow
acceleration is rather complicated \citep*[][]{2014ApJ...783...51C},
which is beyond the scope of this work. Here, we adopt a simplified
model of the outflows
%internal shock jet
%model widely used for Gamma-ray bursts and active galactic nuclei
%\citep*[see, e.g.,][]{1999PhR...314..575P,2001MNRAS.325.1559S}.
with a conical geometry \citep*[see][for the
details]{2010ApJ...724..855C}.

The density of a conical outflow at a distance of $R$ from the black
hole is
\begin{equation}
\rho(R)={\frac {\dot{M}_{\rm w}}{4\pi R^2\gamma_{\rm w}\beta_{\rm
w}c(1-\cos\theta_{\rm w}/2)}}, \label{dens_w}
\end{equation}
where $\dot{M}_{\rm w}$ is the mass loss rate of a pair of outflows,
$\theta_{\rm w}$ is the opening angle of the conical outflows,
$\beta_{\rm w}=v_{\rm w}/c$ is the bulk radial velocity of the
outflows, and
\begin{equation}
\gamma_{\rm w}={\frac {1}{(1-\beta_{\rm w}^2)^{1/2}}}.
\end{equation}
%The mass loss rate $\dot{M}_{\rm w}$ in the outflows is related to
%the Eddington accretion rate $\dot{M}_{\rm Edd}$ with
%\begin{equation}
%\dot{m}_{\rm w}={\frac {\dot{M}_{\rm w}}{\dot{M}_{\rm Edd}}},
%\label{eta_w}
%\end{equation}
%where $\dot{m}_{\rm w}$ is the dimensionless mass loss rate, and
%\begin{displaymath}
%\dot{M}_{\rm Edd}=1.39\times 10^{18}m~{\rm g~s^{-1}},~~~~m={\frac
%{M_{\rm bh}}{M_\odot}}.
%\end{displaymath}
%The electron number density $n_{\rm e}$ is
%\begin{equation}
%n_{\rm e}(r)=2.22\times10^{19}m^{-1}\dot{m}_{\rm w}\gamma_{\rm
%w}^{-1}\beta_{\rm w}^{-1}r^{-2}\left(1-\cos{\frac {\theta_{\rm
%w}}{2}}\right)~{\rm cm}^{-3},\label{n_e}
%\end{equation}
The outflow expanding adiabatically is a good approximation, which
leads to $T\propto R^{-4/3}$ \citep*[see][for the detailed
calculations, but for the outflow with a constant velocity along $R$
in this work]{2010ApJ...724..855C}. The kinetic power of the
outflows can be calculated with
\begin{equation}
P_{\rm kin}=(\gamma_{\rm w}-1)\dot{M}_{\rm w}c^2, \label{p_jet}
\end{equation}
where $r=R/R_{\rm S}$, $R_{\rm S}=2GM_{\rm bh}/c^2$, and $m=M_{\rm
bh}/M_\odot$. Substitute Equation (\ref{p_jet}) into Equation
(\ref{dens_w}), we have
\begin{equation}
n_{\rm e}(r)=1.77\times10^{-20}m^{-2}P_{\rm kin}(\gamma_{\rm
w}-1)^{-1}\gamma_{\rm w}^{-1}\beta_{\rm
w}^{-1}r^{-2}\left(1-\cos{\frac {\theta_{\rm w}}{2}}\right)~{\rm
cm}^{-3}.\label{n_e2}
\end{equation}

As the gas in the outflows is driven from the corona, only a small
fraction of the electrons may be re-accelerated in shocks to a
non-thermal component, and the outflows should contain a dominant
component of thermal gas.
%In principle, the electron temperature of
%the gas in the jets can be calculated based on the internal shock
%model \citep*[][]{1999PhR...314..575P}, if the detailed physics of
%the jets is properly considered. However, such calculations will be
%suffered from poor understand of the detailed physical processes in
%the jets, and the results may still be dependent on the model
%parameters \citep*[][]{1999PhR...314..575P}. In this work, we focus
%on how the jet properties can be constrained by the observed X-ray
%emission lines in 4U~1630$-$47. For simplicity, we adopt the
%electron temperature $T_{\rm e}$ as an input model parameter, and
%$f_{\rm th}$ to describe the thermal electron fraction in the jets,
%which are sufficient for the calculations of the minimal kinetic
%power (mass loss rate) of the jets required to produce the observed
%thermal X-ray lines.
In this work, we assume a power-law $R$-dependent electron
temperature in the outflows,
\begin{equation}
T_{\rm e}(R)=T_{\rm e,in}\left({\frac {R}{R_{\rm
in}}}\right)^{-\xi_{\rm T}}, \label{t_e1}
\end{equation}
where $R_{\rm in}$ is the location of the base of the outflows, and
the thermal electron temperature is $T_{\rm e,in}$ at $R=R_{\rm
in}$. For an adiabatically expanding outflow, the index $\xi_{\rm
T}=4/3$ \citep*[][]{2010ApJ...724..855C}.

\section{Thermal X-ray line emission from the
outflows}\label{therm_line}

The thermal X-ray line emissions from the outflows can be
calculated, when the temperature, density, and metallicity of the
gas, are specified. Using the outflow model described in Sect.
\ref{model}, we can calculate the thermal X-ray line emissions from
the outflows. The total line luminosity emitted from the outflows,
$L_{\rm line}$, can be calculated by integrating over $R$ in the
outflows,
\begin{equation}
L_{\rm line}={\frac {Z}{Z_\odot}}\int^{R_{\rm out}}_{R_{\rm
in}}\epsilon[T_{\rm e}(R)]n_{\rm e,th}^{2}(R) 4\pi R^2
\left(1-\cos{\frac {\theta_{\rm w}}{2}}\right){\rm d}R,\label{eq5}
\end{equation}
where $R_{\rm in}$ is the location of the jet base, $n_{\rm e,th}$
is the density of the thermal electrons, and $Z$ is the metallicity
of the gas. Substitute Equation (\ref{n_e2}) into Equation
(\ref{eq5}), we derive the line luminosity,
\begin{displaymath}
L_{\rm line}=1.01\times 10^{-22}m^{-1}P_{\rm kin}^2(\gamma_{\rm
w}-1)^{-2}\gamma_{\rm w}^{-2}\beta_{\rm w}^{-2}\left(1-\cos{\frac
{\theta_{\rm w}}{2}}\right)^3f_{\rm th}^2\left({\frac
{Z}{Z_\odot}}\right)~~~~~~~~~~~~
\end{displaymath}
\begin{equation}
~~~~~~~~~~~~~~~\times\int^{r_{\rm out}}_{r_{\rm in}}\epsilon[T_{\rm
e}(r)]r^{-2}dr~{\rm erg~s}^{-1}, \label{eq5b}
\end{equation}
where a parameter $f_{\rm th}$ is used to describe the fraction of
the thermal electrons in the outflows, i.e., $n_{\rm e,th}=f_{\rm
th}n_{\rm e}$. The line emissivity $\epsilon$ as a function of
temperature is calculated with the standard software package
Astrophysical Plasma Emission Code (APEC)\citep{s01} assuming the
solar metallicity. The code has been used to calculate the X-ray
emission lines emitted from the accretion disk/corona systems
\citep*[e.g.,][]{1999ApJ...515L..69N,2000ApJ...541..898P,2006ApJ...640..319X,2013ApJ...763...75X}.
The collisional excitation, recombination to excited levels and
dielectronic satellite lines have been included in APEC code
\citep*[see][for the details]{s01}.
%The photo-ionization is neglected in our calculations, which is not
%important in the low-luminosity AGNs \citep*[see discussion
%in][]{dgds04}. {The metallicity of M81 has been studied in many
%works
%\citep*[e.g.,][]{1995AJ....110..620P,2000AJ....119.2745K,2012ApJ...747...15K}.
%In this work, we use the value given by \citet{2012ApJ...747...15K}
%which showed that the logrithmic metallicity of M81 relative to the
%sun as a function of galactocentric distance is
%\begin{equation}
%[Z]=(0.286\pm 0.061)-(0.033\pm 0.009)R/{\rm kpc}.
%\end{equation}
%The corresponding metallicity of M81* relative to the solar
%metallicity should be $Z_{\rm m}\simeq 10^{0.286}\simeq 1.9$.
Assuming ionization equilibrium in the plasma, we can thus calculate
the total line luminosities of the X-ray line emissions with the
structure (e.g., temperature and density) of the outflows.

\section{Results}\label{results}

\begin{figure}
\epsscale{0.7} \plotone{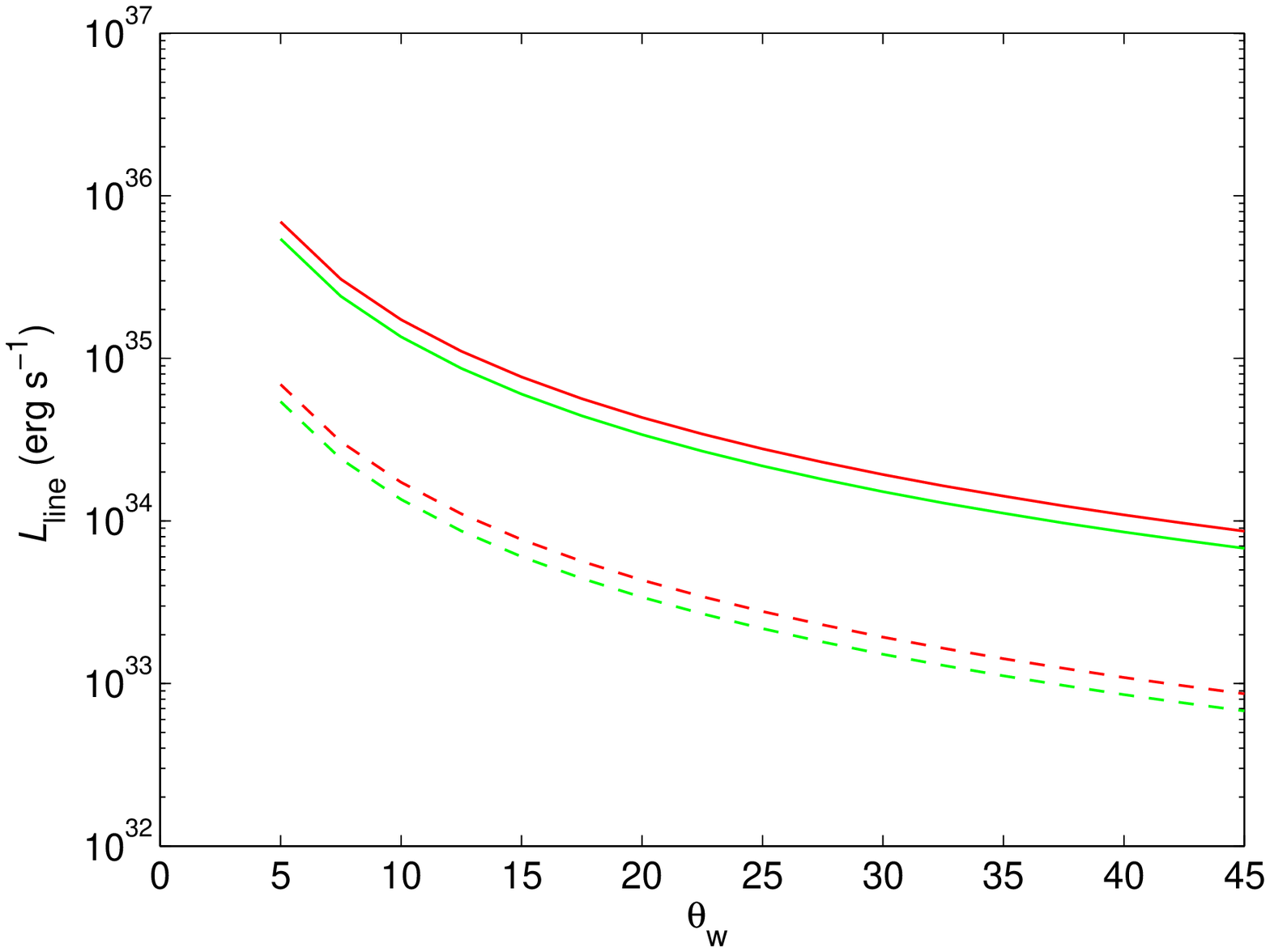} \caption{The luminosities
of the line emitted from the outflows as functions of the opening
angle $\theta_{\rm w}$ of the outflows. The kinetic power of the
outflows $P_{\rm kin}=10^{39}~{\rm erg~s}^{-1}$, and the outflow
velocity $\beta_{\rm w}=0.3$, are adopted in the calculations. The
red lines are the luminosities of the Fe~{\sc xxvi} K$\alpha$
emission line for the outflows, while the green lines are the
Fe~{\sc xxv} K$\alpha$ emission line luminosity. The solid lines
represent the results calculated with $M_{\rm bh}=10M_\odot$, while
the dashed lines are for $M_{\rm bh}=100M_\odot$.
\label{theta_l_line} }
\end{figure}

The structure of the outflow is available when the black hole mass
$M_{\rm bh}$, the kinetic power of the outflows $P_{\rm kin}$, the
outflow velocity $\beta_{\rm w}$, and the opening angle of the
outflow $\theta_{\rm w}$ are specified (see Sect. \ref{model}). The
thermal line luminosity can be calculated with the derived outflow
structure by using the standard software package Astrophysical
Plasma Emission Code (APEC)\citep{s01}. The electron temperature
$T_{\rm e,in}=10^9~{\rm K}$ at the base of the outflow $R_{\rm
in}=5R_{\rm S}$ are adopted in most of the calculations. In this
work, the fraction of the thermal electrons in the outflows $f_{\rm
th}=0.9$ is used, the precise value of which will not affect much on
the main results of this work. The solar metallicity is adopted in
all of the calculations. In Figure \ref{beta_l_line}, we plot the
thermal line luminosities as functions of the outflow velocity
$\beta_{\rm w}$. The dependence of the line luminosity with the
opening angle of the outflow $\theta_{\rm w}$ is plotted in Figure
\ref{theta_l_line}.

\begin{figure}
\epsscale{0.7} \plotone{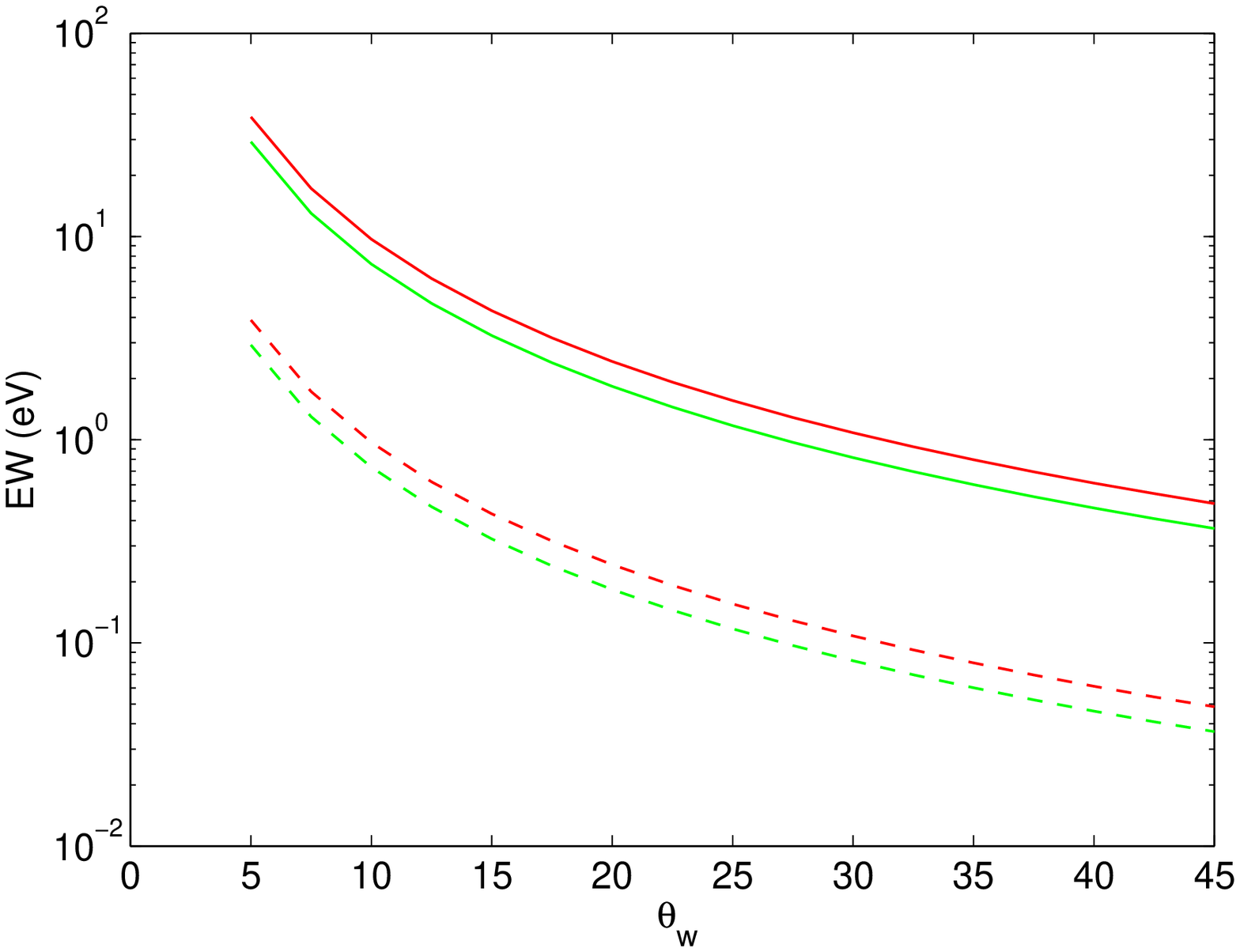} \caption{The equivalent widths
of the line emitted from the outflows as functions of the opening
angle $\theta_{\rm w}$ of the outflows. The accretion power $L_{\rm
acc}=P_{\rm kin}=10^{39}~{\rm erg~s}^{-1}$, and the outflow velocity
$\beta_{\rm w}=0.3$, are adopted in the calculations. The red lines
are the luminosities of the Fe~{\sc xxvi} K$\alpha$ emission line
for the outflows, while the green lines are the Fe~{\sc xxv}
K$\alpha$ emission line luminosity. The solid lines represent the
results calculated with $M_{\rm bh}=10M_\odot$, while the dashed
lines are for $M_{\rm bh}=100M_\odot$. \label{theta_ew}}
\end{figure}

In order to calculate the equivalent width of the X-ray emission
line, we need to know the X-ray continuum spectrum of the ULX. In
this work, a typical template spectrum, of which the photon spectral
index $\Gamma=2$ for the 2$-$10~keV power law spectrum, is adopted.
The fraction of the X-ray luminosity in 2$-$10~keV to the bolometric
luminosity is roughly taken as $0.2$ similar to X-ray binaries
\citep*[e.g.,][]{2013Natur.504..260D}. Assuming the accretion power
$L_{\rm acc}=P_{\rm kin}$, we calculate the equivalent widths of the
lines as functions of $\beta_{\rm w}$ in Figure \ref{beta_ew}. The
equivalent widths varying with the opening angle of the outflow
$\theta_{\rm w}$ is given in Figure \ref{theta_ew}. The thermal line
emission is sensitively dependent of the temperature of the
outflows. In Figure \ref{t0_ew}, we plot the equivalent widths as
functions of the electron temperature of the gas at the base of the
outflows.

\section{Discussion}\label{discussion}

%The thermal line luminosity $L_{\rm line}\propto m^{-1}P_{\rm
%jet}^2$ (see Equation \ref{eq5b}), which indicates that the line
%emission may more easily be detected in the sources with smaller
%black hole masses or/and higher kinetic-power-outflows.

The thermal line emission is proportional to $n_{\rm e}^2$ (see
Equation \ref{eq5}), and $P_{\rm kin}\propto n_{\rm e}$, so the
thermal X-ray line luminosity $L_{\rm line}\propto P_{\rm kin}^2$.
The thermal line luminosity $L_{\rm line}\propto m^{-1}$ (see
Equation \ref{eq5b}), which indicates that the line emission may
more easily be detected in the sources with smaller black hole
masses or/and higher kinetic-power-outflows. The electron density
$n_{\rm e}\propto R^{-2}$ for a conical outflow, i.e., the electron
number density decreases with increasing $R$, and therefore the
thermal line luminosity $L_{\rm line}$ decreases with increasing
$R_{\rm in}$. Similarly, the thermal X-ray line luminosity decreases
with increasing opening angle of the conical outflows for a fixed
kinetic power of the outflows. In order to estimate the observed
equivalent widths of the thermal X-ray emission lines, we use a
template X-ray continuum spectrum of the disk. The continuum
luminosity is assumed to be proportional to the accretion power
$L_{\rm acc}$, so the observed equivalent width, ${\rm EW}\propto
L_{\rm line}/L_{\rm acc}\propto P_{\rm kin}$, for a given ratio
$P_{\rm kin}/L_{\rm acc}$, because $L_{\rm line}\propto m^{-1}$.

\begin{figure}
\epsscale{0.7} \plotone{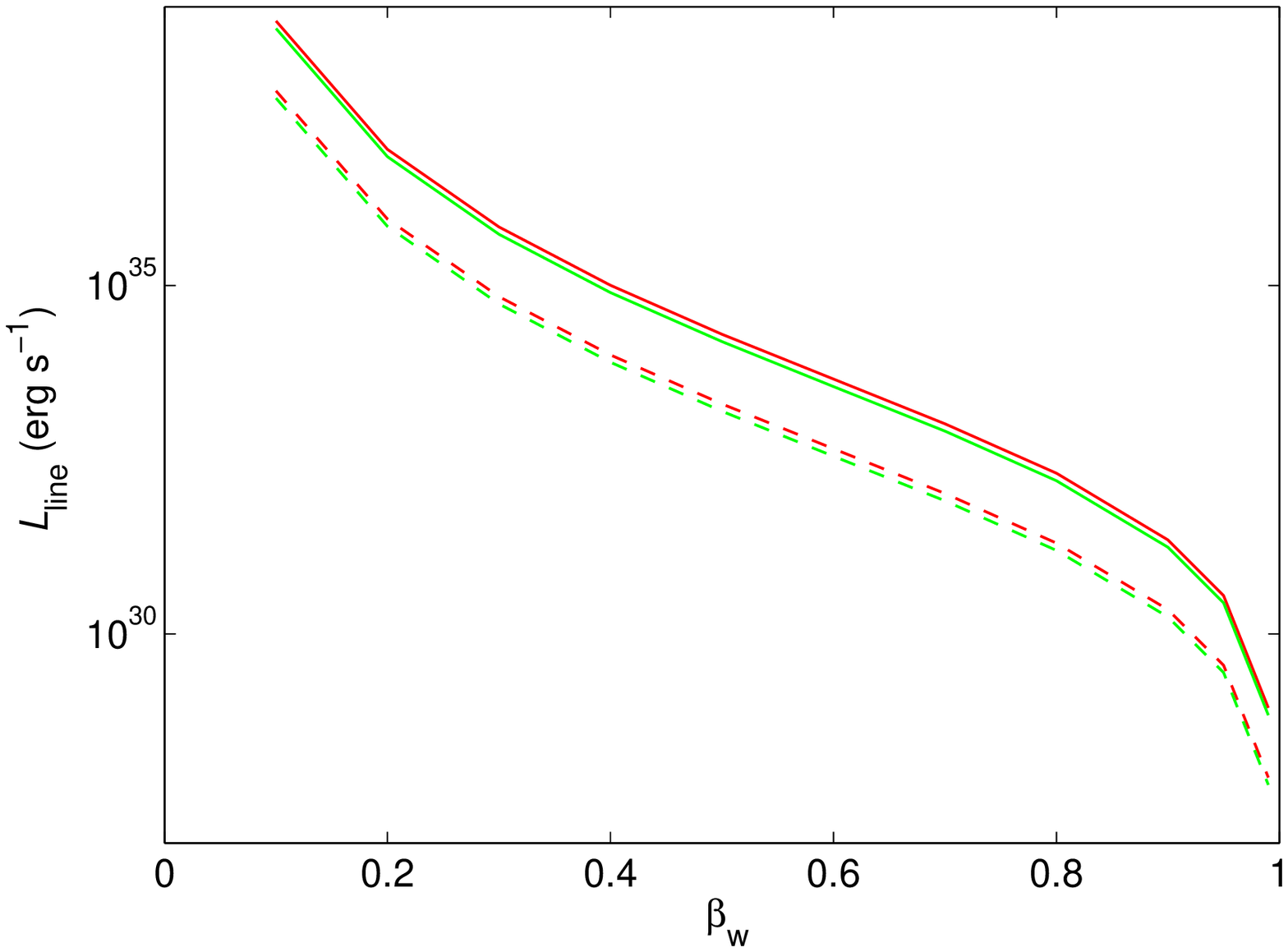} \caption{The luminosities
of the line emitted from the outflows as functions of the outflow
velocity. The kinetic power $P_{\rm kin}=10^{39}~{\rm erg~s}^{-1}$,
and the opening angle $\theta_{\rm w}=5^\circ$ of the outflows are
adopted in the calculations. The red lines are the luminosities of
the Fe~{\sc xxvi} K$\alpha$ emission line for the outflows, while
the green lines are the Fe~{\sc xxv} K$\alpha$ emission line
luminosity. The solid lines represent the results calculated with
$M_{\rm bh}=10M_\odot$, while the dashed lines are for $M_{\rm
bh}=100M_\odot$.\label{beta_l_line} }
\end{figure}

\begin{figure}
\epsscale{0.7} \plotone{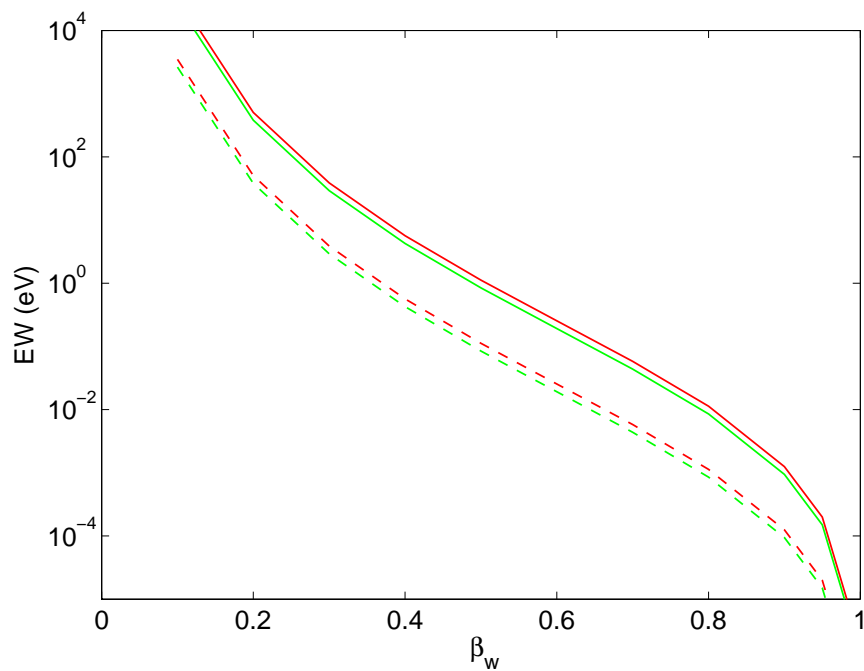} \caption{The equivalent widths
of the line emission as functions of the outflow velocity. The
accretion power $L_{\rm acc}=P_{\rm kin}=10^{39}~{\rm erg~s}^{-1}$,
and the opening angle $\theta_{\rm w}=5^\circ$ are adopted in the
calculations. The red lines are the luminosities of the Fe~{\sc
xxvi} K$\alpha$ emission line for the outflows, while the green
lines are the Fe~{\sc xxv} K$\alpha$ emission line luminosity. The
solid lines represent the results calculated with $M_{\rm
bh}=10M_\odot$, while the dashed lines are for $M_{\rm
bh}=100M_\odot$.\label{beta_ew}}
\end{figure}

\begin{figure}
\epsscale{0.7} \plotone{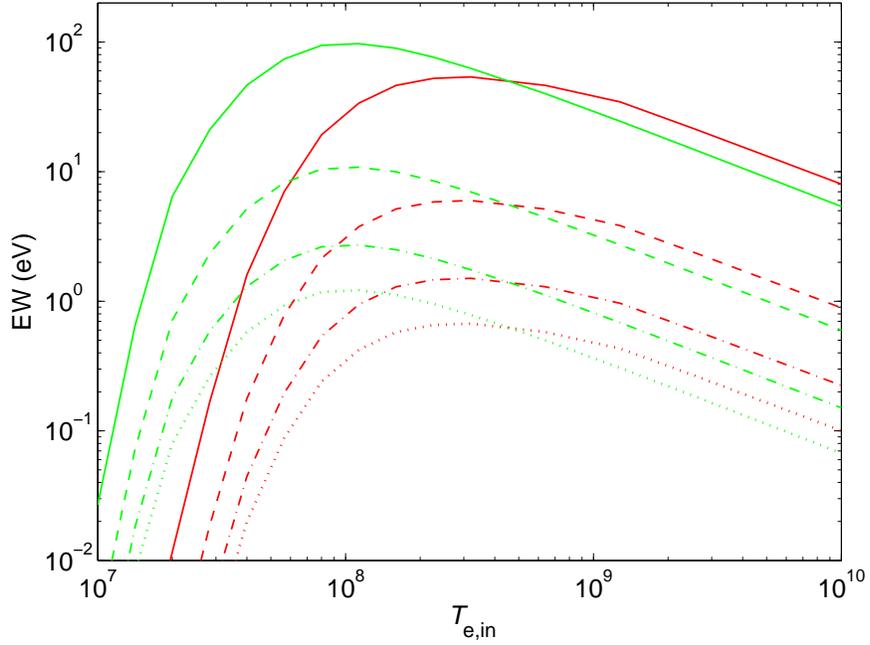} \caption{The equivalent widths of
the line emitted from the outflows as functions of the electron
temperature at the base of the outflow. The black hole mass $M_{\rm
bh}=10M_\odot$, the accretion power $L_{\rm acc}=P_{\rm
kin}=10^{39}~{\rm erg~s}^{-1}$, and the outflow velocity $\beta_{\rm
j}=0.3$, are adopted in the calculations. The red lines are the
luminosities of the Fe~{\sc xxvi} K$\alpha$ emission line for the
outflows, while the green lines are the Fe~{\sc xxv} K$\alpha$
emission line luminosity. The different types of the lines represent
the results calculated with the different opening angle of the
outflow (solid lines: $\theta_{\rm j}=5^\circ$; dashed lines:
$\theta_{\rm j}=15^\circ$; dash-dotted lines: $\theta_{\rm
j}=30^\circ$; dotted lines: $\theta_{\rm j}=45^\circ$).
 \label{t0_ew}}
\end{figure}

The emissivity of the Fe K$\alpha$ line varies with the electron
temperature. We explore how the equivalent widths vary with the
electron temperature $T_{\rm e,in}$ at the base of the outflows (see
Figure \ref{t0_ew}). We find that the equivalent widths are large
when the temperature of the electrons at the base of the outflows is
$\sim 10^{8-9}$~K. The equivalent widths of the iron lines can be
$\sim 1$~eV even for the outflows with an opening angle $\theta_{\rm
w}=45^\circ$ in the ULX containing a stellar mass black hole with
$M_{\rm bh}=10M_\odot$, while it decreases to $\la 0.1$~eV for a
black hole mass with $M_{\rm bh}\ga 100M_\odot$. This means that the
line emission from the outflows sensitively depends on the black
hole mass, because the size of the outflow is proportional to the
black hole mass, and the density of the gas $n_{\rm e}\propto
1/M_{\rm bh}$. {The equivalent widths of the Fe K$\alpha$ lines
depend on outflow velocity sensitively (see Figures
\ref{beta_l_line} and \ref{beta_ew}). The line luminosity $L_{\rm
line}\propto \beta_{\rm w}^{-6}$ for a non-relativistic outflow,
i.e., $\gamma_{\rm w}\rightarrow 1$ in Equation (\ref{eq5b}). It
means that the equivalent widths can be higher for the outflows with
a lower velocity. If the outflow velocity $\beta_{\rm w}=0.2$, the
equivalent widths of the Fe K$\alpha$ lines would be around 10~eV.
The equivalent widths of the soft X-ray atomic lines from the
outflow in the ULX NGC~1313 X-1 are $\sim 0.01-0.17$~keV
\citep*[][]{2015MNRAS.454.3134M}. It was suggested that some ULXs
may contain neutron stars \citep*[][]{2016MNRAS.458L..10K}. Our
model can also be applied for the outflows driven from the accretion
disk surrounding a neutron star, its line luminosity can be several
times higher than that for a stellar black hole case for the same
kinetic power of the outflows, because $L_{\rm line}\propto 1/M$.
The equivalent widths equivalent widths of the Fe K$\alpha$ lines
can be as high as several tens of eV for the ULX containing a
neutron star. }

{We have not considered the line profile in our calculations. The
ion temperature $T{\rm i}\sim 10^{11-12}$K of the corona in the
inner region of the disk, so the ion temperature of the gas at the
base of the outflows $T{\rm i}\la 10^{11-12}$K. The widths of the Fe
K$\alpha$ lines due to thermal broadening are $\la 0.2-0.6$~keV. For
the outflows magnetically driven from the disk, the gas in the
outflows is also rotating \citep*[][]{2010LNP...794..233S}. If the
outflow is viewed edge-on, the line broadening due to rotation of
the gas in the outflow is $\sim 0.25$~keV (for $v_{\rm w}=0.2$c)
assuming the rotating velocity to be an order of magnitude lower
than the radial velocity \citep*[][]{2010LNP...794..233S}. This
means that the rotational broadening of the iron K$\alpha$ emission
line is roughly at the same order of the thermal broadening. }

{The dynamics of the outflows has not been considered in this work.
A simplified model of an outflow with a constant radial velocity is
adopted in this work. We take the outflow velocity of as a model
parameter. The realistic outflow may experience re-acceleration, and
its velocity may vary with the distance from the black hole. The Fe
K$\alpha$ line luminosity for such an outflow can be roughly
estimated by interpolation of the results of the outflows with
constant velocities presented in this work. }

{For the relativistic jets, the line emission may be strongly beamed
if it is viewed at a small angle with line of sight, however, the
line emission decreases significantly with increasing jet velocity
for a fixed kinetic jet power (see Figures \ref{beta_l_line} and
\ref{beta_ew}). In this work, we focus on the slowly moving
outflows, and we have not considered the Doppler beaming effect in
our calculations, which will not affect our main conclusions.  }

In this work, the ratio $P_{\rm kin}/L_{\rm acc}=1$ is adopted in
our calculations, though the values of the ratios are quite unclear
for the ULXs. If it is significantly lower than unity, the thermal
line emission from the outflows can hardly be detected even if the
ULX contains a stellar mass black hole. The ratio of the jet power
to the accretion power for active galactic nuclei (AGN) has been
extensively studied, which shows the jet power is roughly comparable
with the accretion power in most radio-loud AGNs, and the jet power
can be significantly higher than the accretion power in some
luminous radio galaxies
\citep*[e.g.,][]{2009MNRAS.396..984G,2011MNRAS.411.1909F,2011ApJ...727...39M,2014Natur.515..376G}.
Recently, a thin accretion disk with magnetically driven jets is
suggested to explain extremely powerful jets observed in some radio
galaxies \citep*[][]{2014ApJ...788...71L}. {For a conventional thin
disk, almost all the kinetic energy of the rotating gas in the disk
is dissipated in the disk and is radiated out locally. When the
magnetically driven jets are present, a fraction of the kinetic
energy/angular momentum of the disk is removed from the disk by the
jets, and the remainder is dissipated in the disk. An ideal extreme
case is an dissipateless accretion disk can be purely driven by
magnetic jets/outflows, i.e., all the rotating energy of the gas in
the disk is tapped into the jets/outflows.} If most gravitational
energy and angular momentum of the disk is carried away in the jets,
and therefore the jet power can be much higher than the accretion
power \citep*[see][for the details]{2014ApJ...788...71L}. We believe
such mechanism may also work in the accretion-outflow systems of the
ULXs, if the outflow is driven with the Blandford-Payne mechanism.

The solar metallicity $Z=Z_\odot$ is adopted in our calculations
(see Section \ref{results}). We note that the X-ray thermal line
emission is proportional to the metallicity (see Equation
\ref{eq5}). It is shown that the dispersion of the metallicity of
the stars in our galaxy is $\sim 0.5$~dex
\citep*[e.g.,][]{2003AJ....125.1397C,2004IAUS..217..194C,2008IAUS..248..433C,2015RAA....15.1240H}.
The metallicity of nearby galaxies does not deviate much from the
solar value
\citep*[e.g.,][]{1995AJ....110..620P,2000AJ....119.2745K,2012ApJ...747...15K,2016A&A...588A..91M}.
This means that our results may not be altered much by the
metallicity.

In summary, our results suggest that the thermal X-ray Fe line
emission should be preferentially be detected in the ULXs with high
kinetic power outflows from the accretion disks surrounding stellar
mass black holes/neutron stars, or the X-ray atomic features of the
outflows may strongly suggest a stellar mass black hole in the ULX
\citep*[][]{2015MNRAS.454.3134M,2016.1604.08593}.

\acknowledgments We thank the referee for his/her very helpful
comments/suggestions on the manuscript. This work is supported by
the NSFC (grants 11233006, and 11220101002). the Strategic Priority
Research Program ¡°the Emergence of Cosmological Structures¡± of the
CAS (grant No. XDB09000000), and Shanghai Municipality.

{}

%fig 1

%fig 2

%fig 3

%fig 4

%fig 5

\begin{thebibliography}{}



\bibitem[Blandford
\& Payne(1982)]{1982MNRAS.199..883B} Blandford, R.~D., \& Payne,
D.~G.\ 1982, \mnras, 199, 883

%\bibitem[Blandford
%\& Znajek(1977)]{1977MNRAS.179..433B} Blandford, R.~D., \& Znajek,
%R.~L.\ 1977, \mnras, 179, 433

\bibitem[Cao(2004)]{2004ApJ...613..716C} Cao, X.\ 2004, \apj, 613, 716

\bibitem[Cao(2009)]{2009MNRAS.394..207C} Cao, X.\ 2009, \mnras, 394, 207

\bibitem[Cao(2010)]{2010ApJ...724..855C} Cao, X.\ 2010, \apj, 724, 855

\bibitem[Cao(2014)]{2014ApJ...783...51C} Cao, X.\ 2014, \apj, 783, 51

\bibitem[Chen et al.(2003)]{2003AJ....125.1397C} Chen, L., Hou, J.~L.,
\& Wang, J.~J.\ 2003, \aj, 125, 1397

\bibitem[Chen et al.(2004)]{2004IAUS..217..194C} Chen, L., Hou, J.~L.,
\& Wang, J.~J.\ 2004, Recycling Intergalactic and Interstellar
Matter, 217, 194

\bibitem[Chen et al.(2008)]{2008IAUS..248..433C} Chen, L., Hou, J.~L.,
Zhao, J.~L., \& de Grijs, R.\ 2008, IAU Symposium, 248, 433

\bibitem[Cui et al.(2000)]{2000ApJ...529..952C} Cui, W., Chen, W.,
\& Zhang, S.~N.\ 2000, \apj, 529, 952

\bibitem[D{\'{\i}}az Trigo et al.(2013)]{2013Natur.504..260D} D{\'{\i}}az
Trigo, M., Miller-Jones, J.~C.~A., Migliari, S., Broderick, J.~W.,
\& Tzioumis, T.\ 2013, \nat, 504, 260

\bibitem[Fabrika(2004)]{2004ASPRv..12....1F} Fabrika, S.\ 2004, Astrophysics and Space Physics Reviews, 12, 1

\bibitem[Fabrika \& Mescheryakov(2001)]{2001IAUS..205..268F} Fabrika, S., \& Mescheryakov, A.\ 2001, Galaxies and their Constituents at the Highest Angular Resolutions, 205, 268

\bibitem[Fabrika et al.(2016)]{2016arXiv160105971F} Fabrika, S., Vinokurov, A., \& Atapin, K.\ 2016, arXiv:1601.05971

\bibitem[Fernandes et al.(2011)]{2011MNRAS.411.1909F} Fernandes, C.~A.~C.,
Jarvis, M.~J., Rawlings, S., et al.\ 2011, \mnras, 411, 1909

\bibitem[Ghisellini et al.(2014)]{2014Natur.515..376G} Ghisellini, G.,
Tavecchio, F., Maraschi, L., Celotti, A., \& Sbarrato, T.\ 2014,
\nat, 515, 376

\bibitem[Gu et al.(2009)]{2009MNRAS.396..984G} Gu, M., Cao, X.,
\& Jiang, D.~R.\ 2009, \mnras, 396, 984

%\bibitem[Hjellming et al.(1999)]{1999ApJ...514..383H} Hjellming, R.~M.,
%Rupen, M.~P., Mioduszewski, A.~J., et al.\ 1999, \apj, 514, 383

\bibitem[Huang et al.(2015)]{2015RAA....15.1240H} Huang, Y., Liu, X.-W.,
Zhang, H.-W., et al.\ 2015, Research in Astronomy and Astrophysics,
15, 1240

\bibitem[King \& Lasota(2016)]{2016MNRAS.458L..10K} King, A., \& Lasota, J.-P.\ 2016, \mnras, 458, L10

\bibitem[Kong et al.(2000)]{2000AJ....119.2745K} Kong, X., Zhou, X., Chen, J., et al.\ 2000, \aj, 119, 2745

\bibitem[Kotani et al.(1994)]{1994PASJ...46L.147K} Kotani, T., Kawai, N., Aoki, T., et al.\ 1994, \pasj, 46, L147

\bibitem[Kotani et al.(1996)]{1996PASJ...48..619K} Kotani, T., Kawai, N., Matsuoka, M., \& Brinkmann, W.\ 1996, \pasj, 48, 619

\bibitem[Kudritzki et al.(2012)]{2012ApJ...747...15K} Kudritzki, R.-P., Urbaneja, M.~A., Gazak, Z., et al.\ 2012, \apj, 747, 15


\bibitem[Li(2014)]{2014ApJ...788...71L} Li, S.-L.\ 2014, \apj, 788, 71

\bibitem[Liu et al.(2003)]{2003ApJ...587..571L} Liu, B.~F., Mineshige, S.,
\& Ohsuga, K.\ 2003, \apj, 587, 571

\bibitem[Liu et al.(2013)]{2013Natur.503..500L} Liu, J.-F., Bregman, J.~N., Bai, Y., Justham, S., \& Crowther, P.\ 2013, \nat, 503, 500

\bibitem[Magrini et al.(2016)]{2016A&A...588A..91M} Magrini, L., Coccato, L., Stanghellini, L., Casasola, V., \& Galli, D.\ 2016, \aap, 588, A91

\bibitem[Marshall et al.(2013)]{2013ApJ...775...75M} Marshall, H.~L., Canizares, C.~R., Hillwig, T., et al.\ 2013, \apj, 775, 75

\bibitem[McNamara et al.(2011)]{2011ApJ...727...39M} McNamara, B.~R.,
Rohanizadegan, M., \& Nulsen, P.~E.~J.\ 2011, \apj, 727, 39

\bibitem[Merloni
\& Fabian(2002)]{2002MNRAS.332..165M} Merloni, A., \& Fabian, A.~C.\
2002, \mnras, 332, 165

\bibitem[Middleton et al.(2015)]{2015MNRAS.454.3134M} Middleton, M.~J., Walton, D.~J., Fabian, A., et al.\ 2015, \mnras, 454, 3134

\bibitem[Motch et al.(2014)]{2014Natur.514..198M} Motch, C., Pakull, M.~W., Soria, R., Gris{\'e}, F., \& Pietrzy{\'n}ski, G.\ 2014, \nat, 514, 198

\bibitem[Narayan \& Raymond(1999)]{1999ApJ...515L..69N} Narayan, R., \& Raymond, J.\ 1999, \apjl, 515, L69

\bibitem[Neilsen et al.(2014)]{2014ApJ...784L...5N} Neilsen, J., Coriat,
M., Fender, R., et al.\ 2014, \apjl, 784, L5

%\bibitem[Piran(1999)]{1999PhR...314..575P} Piran, T.\ 1999, \physrep, 314,
%575

\bibitem[Perelmuter et al.(1995)]{1995AJ....110..620P} Perelmuter, J.-M., Brodie, J.~P., \& Huchra, J.~P.\ 1995, \aj, 110, 620

\bibitem[Perna et al.(2000)]{2000ApJ...541..898P} Perna, R., Raymond, J., \& Narayan, R.\ 2000, \apj, 541, 898

%\bibitem[Pinto et al.(2016)]{2016.1604.08593} Pinto, C., Middleton, M.~J., \&
%Fabian, A.~C.\ 2016, \nat, astro-ph. 1604.08593

\bibitem[Pinto et al.(2016)]{2016.1604.08593} Pinto, C., Middleton, M.~J., \& Fabian, A.~C.\ 2016, \nat, 533, 64


\bibitem[Smith et al.(2001)]{s01}
  Smith, R. K., Brickhouse, N. S., Liedahl, D. A., Raymond, J. C. 2001, \apj, 556, L91

\bibitem[Spruit(2010)]{2010LNP...794..233S} Spruit, H.~C.\ 2010, Lecture Notes in Physics, Berlin Springer Verlag, 794, 233


%\bibitem[Spada et al.(2001)]{2001MNRAS.325.1559S} Spada, M., Ghisellini,
%G., Lazzati, D., \& Celotti, A.\ 2001, \mnras, 325, 1559

%\bibitem[Tchekhovskoy et al.(2011)]{2011MNRAS.418L..79T} Tchekhovskoy, A.,
%Narayan, R., \& McKinney, J.~C.\ 2011, \mnras, 418, L79

\bibitem[Watson et al.(1986)]{1986MNRAS.222..261W} Watson, M.~G., Stewart, G.~C., King, A.~R., \& Brinkmann, W.\ 1986, \mnras, 222, 261

\bibitem[Wu et al.(2013)]{2013ApJ...770...31W} Wu, Q., Cao, X., Ho, L.~C.,
\& Wang, D.-X.\ 2013, \apj, 770, 31

%\bibitem[Wu et al.(2007)]{2007ApJ...669...96W} Wu, Q., Yuan, F.,
%\& Cao, X.\ 2007, \apj, 669, 96

\bibitem[Xu(2013)]{2013ApJ...763...75X} Xu, Y.-D.\ 2013, \apj, 763, 75

\bibitem[Xu et al.(2006)]{2006ApJ...640..319X} Xu, Y.-D., Narayan, R.,
Quataert, E., Yuan, F., \& Baganoff, F.~K.\ 2006, \apj, 640, 319

%\bibitem[Yuan et al.(2005)]{2005ApJ...620..905Y} Yuan, F., Cui, W.,
%\& Narayan, R.\ 2005, \apj, 620, 905

\bibitem[Zdziarski et al.(2011)]{2011MNRAS.416.1324Z} Zdziarski, A.~A.,
Skinner, G.~K., Pooley, G.~G., \& Lubi{\'n}ski, P.\ 2011, \mnras,
416, 1324

\end{thebibliography}
\end{document}